\documentclass[preprint]{aastex}
\usepackage{graphicx}
\usepackage{endfloat}
\usepackage{amsmath}
\usepackage{mathtools}

\shorttitle{Trajectory analysis for the nucleus and dust of comet C/2013~A1 (Siding Spring)}
\shortauthors{Farnocchia et al.}
\slugcomment{Submitted}

\begin{document}

\title{Trajectory analysis for the nucleus and dust of comet C/2013~A1 (Siding Spring)}

\author{
Davide Farnocchia\altaffilmark{*}\altaffilmark{a}, 
Steven R. Chesley\altaffilmark{a},
Paul W. Chodas\altaffilmark{a},
Pasquale Tricarico\altaffilmark{b},
Michael S.~P. Kelley\altaffilmark{c},
Tony L. Farnham\altaffilmark{c}
}

\altaffiltext{*}{\tt Davide.Farnocchia@jpl.nasa.gov}

\altaffiltext{a}{Jet Propulsion Laboratory, California Institute of
  Technology, Pasadena, CA 91109, USA} 
\altaffiltext{b}{Planetary Science Institute, Tucson, AZ 85719, USA}
\altaffiltext{c}{Department of Astronomy, University of Maryland,
  College Park, MD 20742, USA}

\begin{abstract}
  Comet C/2013~A1 (siding Spring) will experience a high velocity
  encounter with Mars on October 19, 2014 at a distance of 135,000 km
  $\pm$ 5000 km from the planet center. We present a comprehensive
  analysis of the trajectory of both the comet nucleus and the dust
  tail. The nucleus of C/2013~A1 cannot impact on Mars even in the
  case of unexpectedly large nongravitational
  perturbations. Furthermore, we compute the required ejection
  velocities for the dust grains of the tail to reach Mars as a
  function of particle radius and density and heliocentric distance of
  the ejection. A comparison between our results and the most current
  modeling of the ejection velocities suggests that impacts are
  possible only for millimeter to centimeter size particles released
  more than 13 au from the Sun. However, this level of cometary
  activity that far from the Sun is considered extremely unlikely.
  The arrival time of these particles spans a 20-minute time interval
  centered at October 19, 2014 at 20:09 TDB, i.e., around the time
  that Mars crosses the orbital plane of C/2013~A1. Ejection
  velocities larger than currently estimated by a factor $>2$ would
  allow impacts for smaller particles ejected as close as 3 au from
  the Sun. These particles would reach Mars from 43 to 130 min after
  the nominal close approach epoch of the purely gravitational
  trajectory of the nucleus.
\end{abstract}


\section{Introduction}\label{sec:intro}
Comet C/2013~A1 (Siding Spring) was discovered on January 2013 at the
Siding Spring observatory \citep{McNaught13}. Shortly after discovery
it was clear that C/2013~A1 was headed for a close encounter with Mars
on October 19, 2014. C/2013~A1 is on a near parabolic retrograde orbit
and will have a high relative velocity with respect to Mars of about 56 km/s
during the close approach.

If the comet has no significant nongravitational perturbations, the
trajectory of the nucleus consistent with the present set of
astrometric observations rules out an impact on Mars. However, comet
orbits are generally difficult to predict. As the comet gets closer to
the Sun cometary activity can result in significant nongravitational
perturbations \citep{Marsden73} that in turn can lead to significant
deviations from the purely gravitational (``ballistic'')
trajectory. In the case of C/2013~A1, cometary activity was already
visible in the discovery observations, when the comet was at more than
7 au from the Sun.

Beside the effect of nongravitational perturbations, dust grains in
the tail of the comet could reach Mars and possibly damage spacecrafts
orbiting Mars, i.e., NASA's Mars Reconnaissance Orbiter, NASA's Mars
Odyssey, ESA's Mars Express, NASA's MAVEN, and ISRO's
MOM. \citet{Vaubaillon14} and \citet{Moorhead14} show that dust grains
can reach Mars if they are ejected from the nucleus with a
sufficiently high velocity.

The modeling of the ejection velocity is in continuous evolution. As
the comet gets closer to the inner solar system we have additional
observation that provide constraints to the ejection velocities of
dust grains. In particular, by making use of observations from
HST/WFC3, Swift/UVOT, and WISE, \citet{Farnham14} and
\citet{Tricarico14} find ejection velocities lower than those derived
by \citet{Vaubaillon14} and \citet{Moorhead14}, thus significantly reducing the hazard due to
dust grains in the comet tail.

In this paper we study the trajectory of C/2013~A1's nucleus,
including the contribution of nongravitational perturbations.  We also
present an analysis of the required ejection velocities for the dust
grains to reach Mars. This analysis can be used as a reference as the
understanding and the modeling of the dust grain ejection velocities
evolve.

\section{Ballistic trajectory}
We examined all available ground-based optical astrometry (Right
Ascension and Declination angular pairs) as of March 15, 2014. To
remove biased contributions from individual observatories we
conservatively excluded from the orbital fit batches of more than four
observations in the same night with mean residual larger than
0.5\arcsec, and batches of three or four observations showing mean
residual larger than 1\arcsec. We also adopted the outlier rejection
scheme of \citet{Carpino03} with $\chi_{rej} = 2$. To the remaining
597 optical observations we applied the standard one arcsecond
data-weights used for comet astrometry. Figure 1 shows the residuals
of C/2013~A1's observations against our new orbit solution (JPL
solution 46).

\begin{figure}[t]
\begin{center}
\includegraphics[width=5in]{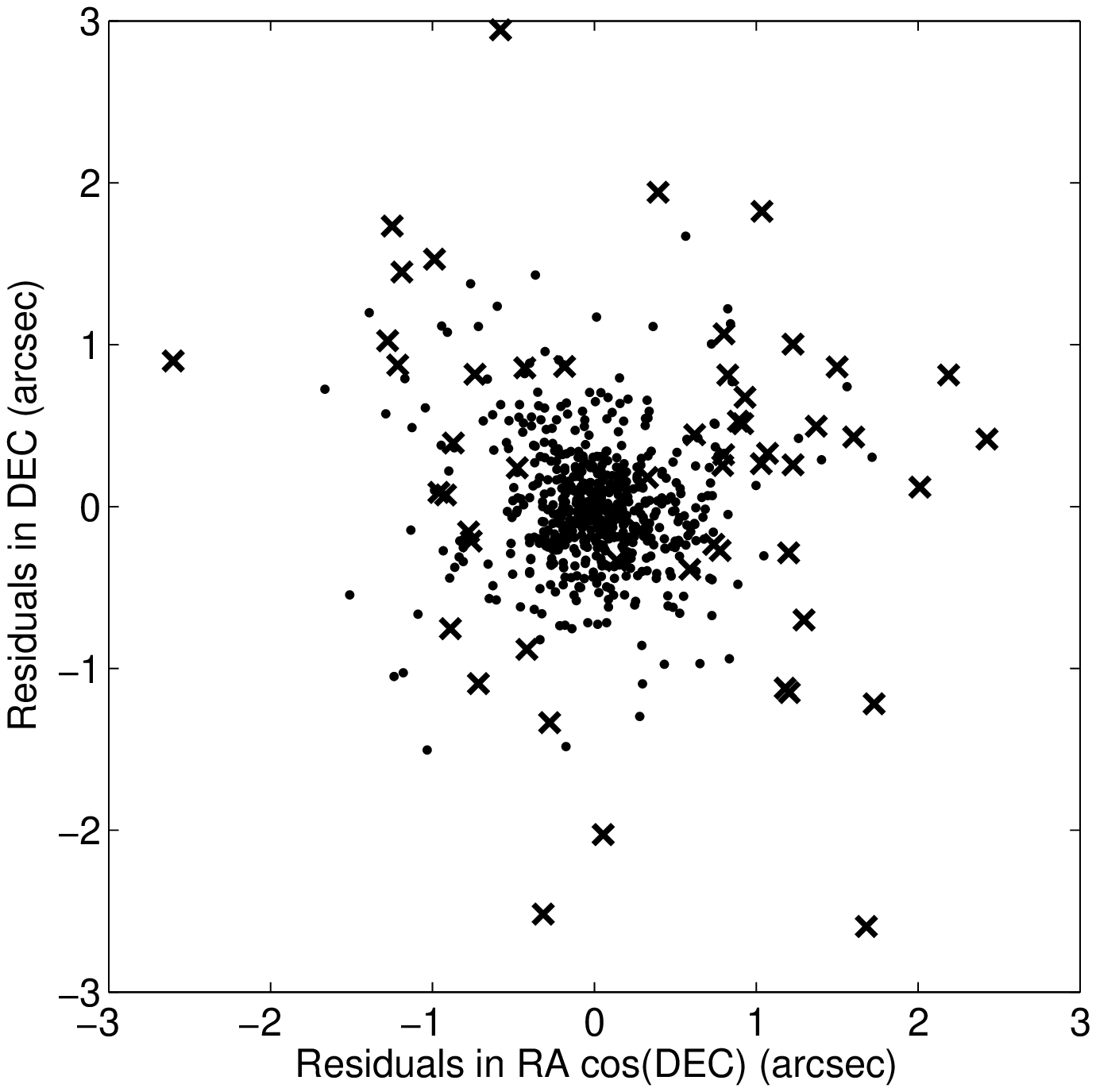}
\caption{Scatter plot of the astrometric residuals in Right Ascension
  and Declination with respect to JPL solution 46. Crosses correspond
  to rejected observations, while dots correspond to the observations
  included in the fit.}
\label{fig:res}
\end{center}
\end{figure}

Our force model included solar and planetary perturbations based on
JPL's planetary ephemerides
DE431\footnote{http://ssd.jpl.nasa.gov/?ephemerides}, the
gravitational attraction due to the 16 most massive bodies in the main
asteroid belt, and the Sun relativistic term. No significant
nongravitational forces were evident in the astrometric data and so
the corresponding JPL orbit solution is ballistic, identified as
number 46. Table~\ref{tab:orbit} contains the orbital elements of the
computed solution.

\begin{table}[ht]
  \caption{J2000 heliocentric ecliptic orbital parameters of JPL orbit
    solution 46. Numbers in parentheses indicate the $1\sigma$ formal
    uncertainties of the corresponding (last two) digits in the
    parameter value.\label{tab:orbit}}
\begin{center}
\begin{tabular}{lc}
\hline
Epoch (TDB) & 2013 Aug 1.0\\
Eccentricity & 1.0006045(61)\\
Perihelion distance (au) & 1.3990370(73)\\
Time of perihelion passage (TDB) & 2014 Oct 25.3868(14)\\
Longitude of node ($^\circ$) & 300.974337(84)\\
Argument of perihelion ($^\circ$) & 2.43550(33)\\
Inclination ($^\circ$) & 129.026659(32)\\
\hline
\end{tabular}
\end{center}
\end{table}

Table~\ref{tab:closeapp} provides information on the close encounter
between C/2013~A1 and Mars. C/2013~A1 passes through the orbital plane
of Mars 69 minutes before the close approach epoch, while Mars passes
through the orbital plane of C/2013 A1 99 minutes after the close approach. The
Minimum Orbit Intersection Distance (MOID) is the minimum distance
between the orbit of the comet and the orbit of Mars
\citep[MOID,][]{Gronchi07}.  The MOID points on the two orbits are not
on the line of nodes. Mars arrives at the minimum distance point 101
min after the close approach epoch, while C/2013~A1 arrives at the
minimum distance point 70 min before the close approach, which means
that the comet is 171 min early for the minimum distance encounter.

\begin{table}[ht]
  \caption{Close approach data for JPL orbit solution 46.\label{tab:closeapp}}
\begin{center}
\begin{tabular}{lc}
\hline
Close approach epoch ($\pm 3\sigma$) & 2014 Oct 19 18:30 (TDB) $\pm$ 3 min\\
Close approach distance ($\pm 3\sigma$) & 134,680 km $\pm$ 4520 km\\
Asymptotic relative velocity ($v_\infty$) & 55.96 km/s\\
MOID & 27,414 km\\
Node crossing distance & 27,563 km\\
Mars's arrival at line of nodes & 2014 Oct 19 20:09 (TDB)\\
C/2013~A1's arrival at line of nodes & 2014 Oct 19 17:21
(TDB)\\
Mars's arrival at MOID & 2014 Oct 19 20:11 (TDB)\\
C/2013~A1's arrival at MOID & 2014 Oct 19 17:20 (TDB)\\
\hline
\end{tabular}
\end{center}
\end{table}

A standard tool to analyze planetary encounters is the $b$-plane
\citep{Kizner61,Valsecchi03}, defined as the plane passing through the
center of mass of the planet and normal to the inbound hyperbolic
approach asymptote. The coordinates on the $b$-plane described in
\citet{Valsecchi03} are oriented such that the projected heliocentric
velocity of the planet is along $-\zeta$. Therefore, $\zeta$ varies
with the time of arrival, i.e., a positive $\zeta$ means that the
comet arrives late at the encounter while a negative $\zeta$ means
that the comet arrives early. On the other hand $\xi$ is related to
the MOID. The $b$-plane is used on a daily basis for asteroid close
approaches to the Earth and computing the corresponding impact
probabilities \citep{Milani05}.

\begin{figure}[t]
\begin{center}
\includegraphics[width=3in]{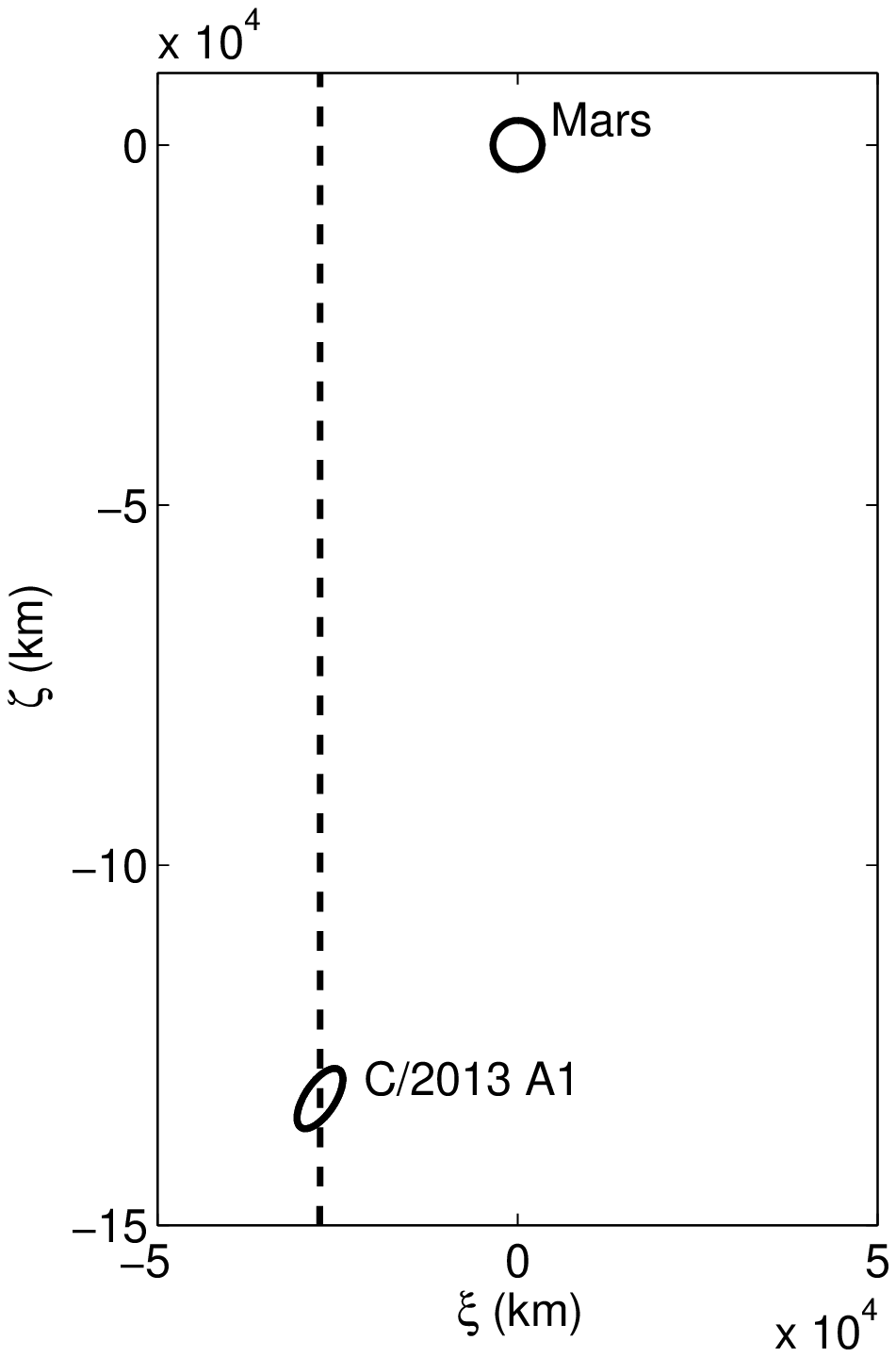}
\caption{Projection of the 3$\sigma$ uncertainty of JPL solution 46 on
  the October 2014 $b$-plane. The nominal prediction for the $b$-plane
  coordinates is $(\xi, \zeta) = (-27,445, -132,407)$ km. The dashed
  line represents the projection of the orbit of C/2013~A1 on the
  $b$-plane. The minimum distance between the orbits of Mars and
  C/2013~A1 is $\sim$27,400 km.}
\label{fig:b_plane}
\end{center}
\end{figure}

Figure~\ref{fig:b_plane} shows the projection of the 3$\sigma$
uncertainty ellipsoid of JPL solution 46 on the $b$-plane. The
projection of the velocity of Mars on this plane is oriented as
$-\zeta$, while the Mars-to-Sun vector projection is on the left side,
at a counterclockwise angle of 186$^\circ$ with respect to the $\xi$
axis. The negative $\zeta$ coordinate of the center of the ellipse
corresponds to the 171 min time shift between Mars and C/2013~A1.

\section{Nongravitational perturbations}\label{sec:nongravs}
Comet trajectories can be significantly affected by nongravitational
perturbations due to cometary outgassing. We use the \citet{Marsden73}
comet nongravitational model:
\begin{equation}\label{eq:nongravs}
\mathbf a_{NG} = g(r) (A_1\hat{\mathbf r} + A_2\hat{\mathbf t} + A_3 \hat{\mathbf n})
\end{equation}
where $g(r)$ is a known function of the heliocentric distance $r$, and
$A_i$ are free parameters that give the nongravitational acceleration
at 1 au in the radial-transverse-normal reference frame defined by
$\hat{\mathbf r}$, $\hat{\mathbf t}$, $\hat{\mathbf n}$.

\begin{figure}[t]
\begin{center}
\includegraphics[width=4.5in]{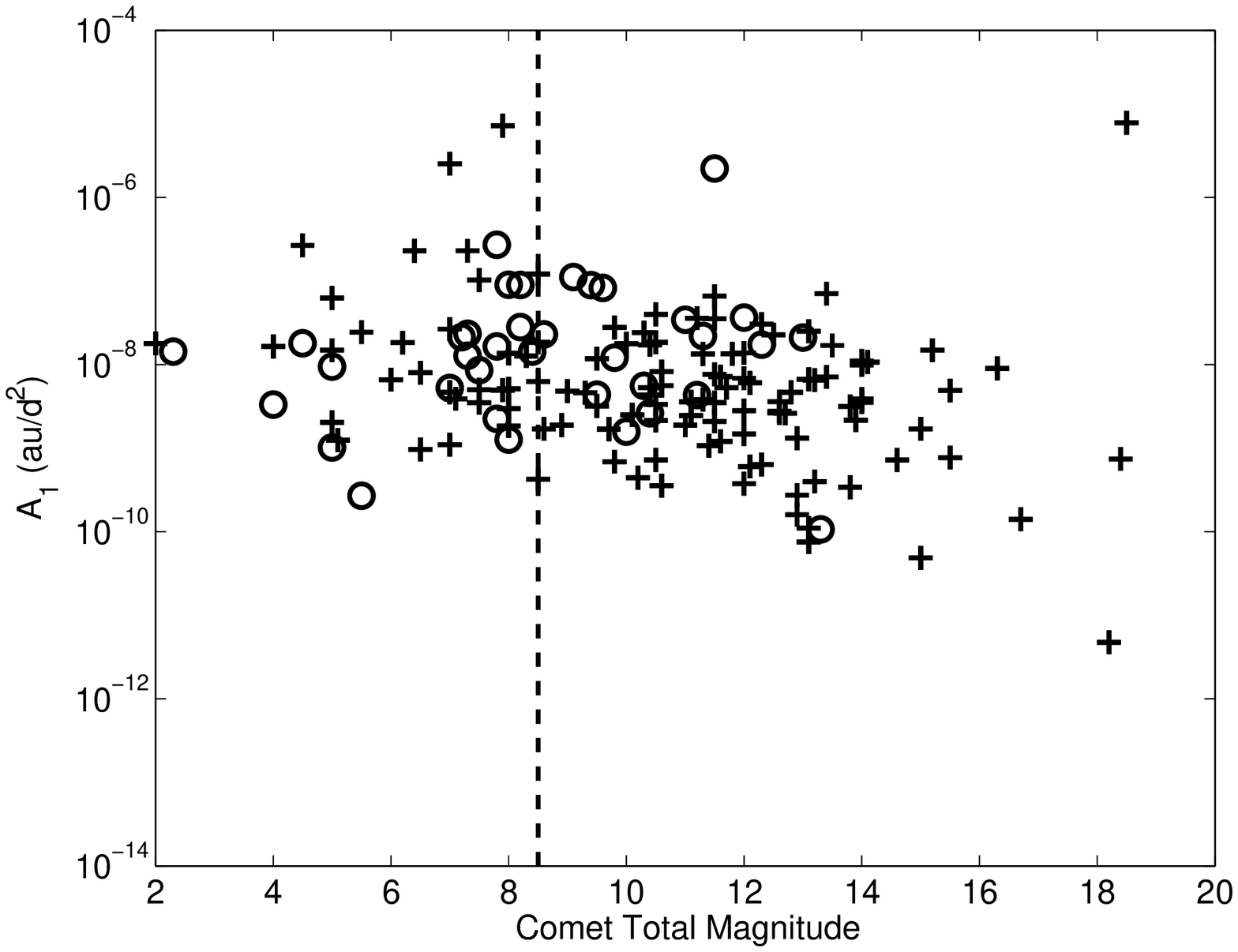}
\includegraphics[width=4.5in]{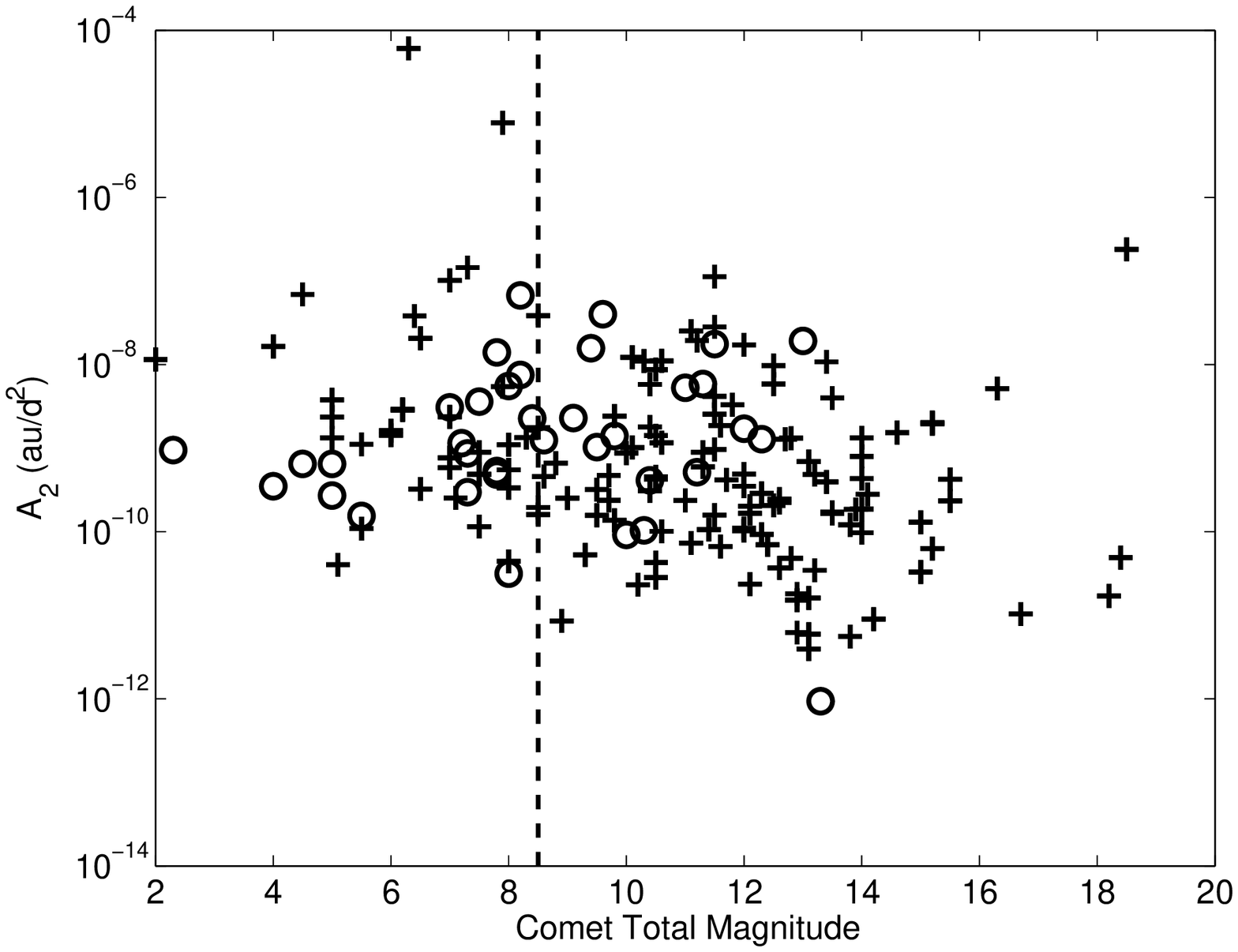}
\caption{Estimated nongravitational parameters $A_1$ and $A_2$ for the
  comets in the catalog. $A_2$ is reported in absolute value. Circles
  correspond to comets with a period larger than 60 yr or an
  eccentricity larger than 0.9. Crosses are for all other comets. The
  dashed line corresponds to the total magnitude of C/2013~A1.}
\label{fig:cat_ng}
\end{center}
\end{figure}

\begin{figure}[t]
\begin{center}
\includegraphics[width=4.5in]{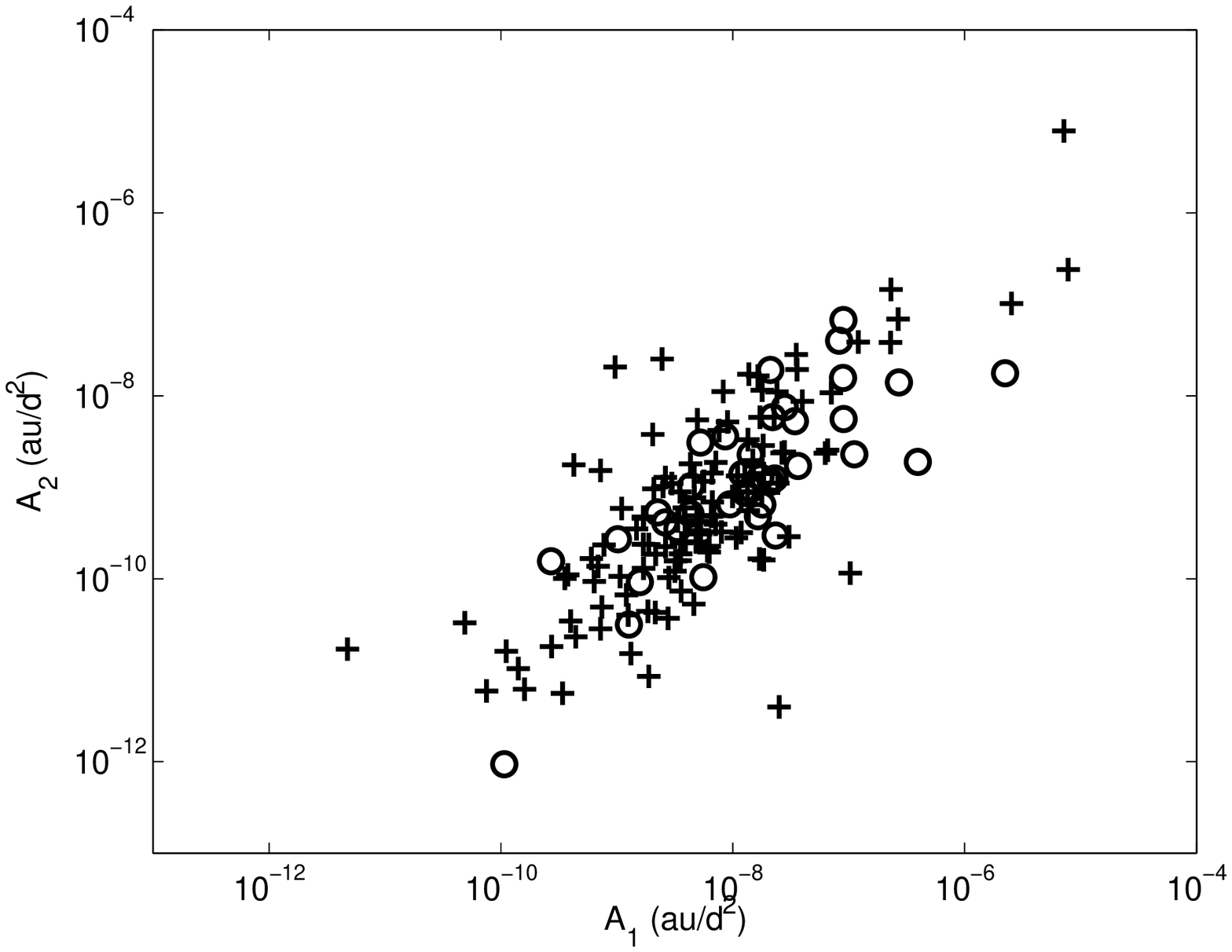}
\includegraphics[width=4.5in]{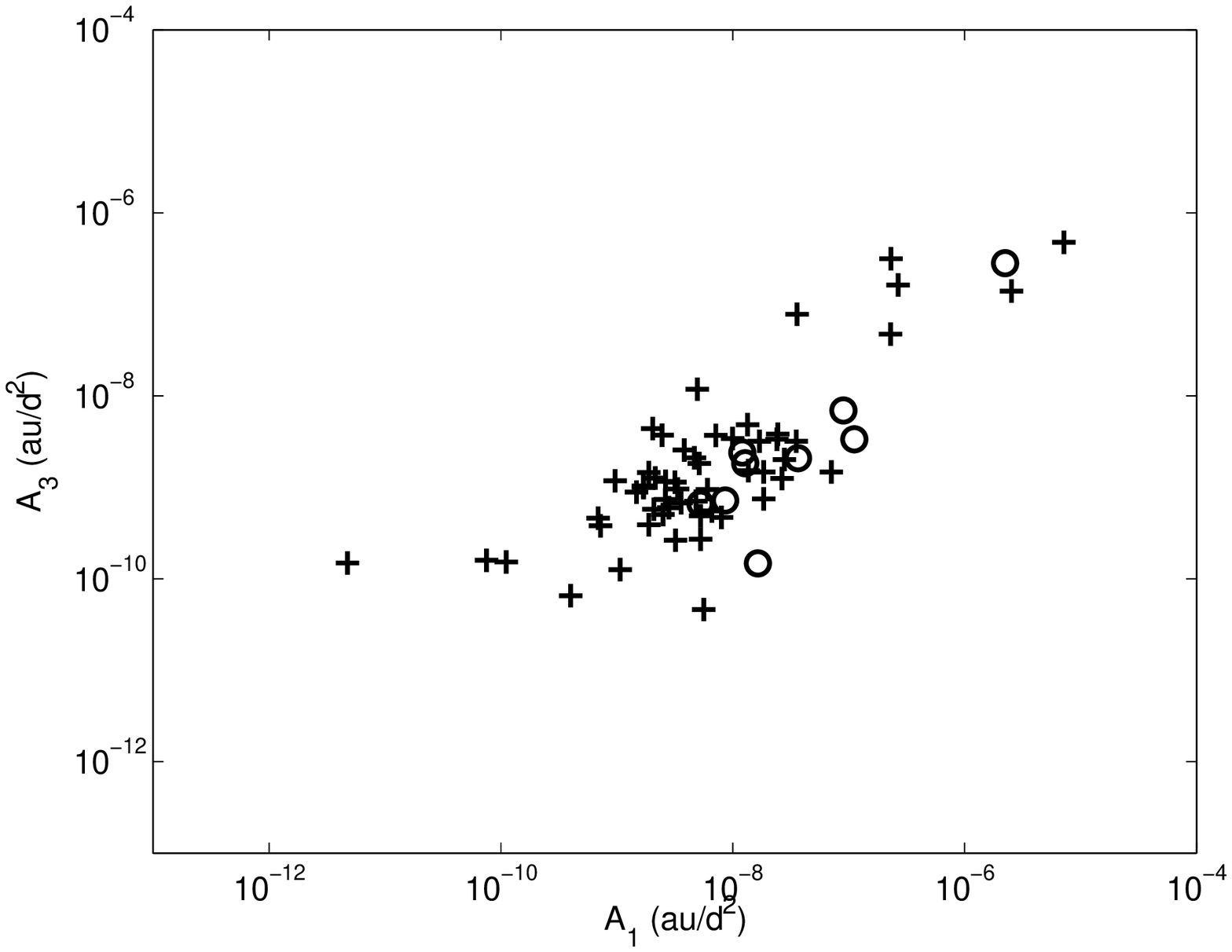}
\caption{Scatter plots for nongravitational parameters $A_1$, $A_2$,
  and $A_3$. $A_2$ and $A_3$ are reported in absolute value. Circles
  correspond to comets with a period larger than 60 yr or an
  eccentricity larger than 0.9. Crosses are for all other comets.}
\label{fig:ng_corr}
\end{center}
\end{figure}

The observational dataset available for C/2013~A1 does not allow us to
estimate the nongravitational parameters $A_i$. Still,
nongravitational accelerations could cause statistically significant
deviations at the close approach epoch. To deal with this problem, we
analyzed the properties of known nongravitational parameters in the
comet catalog. Figure~\ref{fig:cat_ng} shows the known $A_1$ and $A_2$
in the catalog. $A_3$ values have an order of magnitude similar to
that of $A_2$. Figure~\ref{fig:ng_corr} contains scatter plots of
nongravitational parameters showing the correlation between these
parameters. For comets with an orbit similar to that of C/2013~A1,
i.e., with large orbital period ($>$ 60 yr) and high eccentricity ($>$
0.9), values of $A_1$ are on average $\sim 10^{-8}$ au/d$^2$, but they
can be as large as $\sim 10^{-6}$. $A_2$ and $A_3$ are generally one
order of magnitude smaller, i.e., on average they are $\sim 10^{-9}$
au/d$^2$ but can be as large as $\sim 10^{-7}$ au/d$^2$. We can see
that $A_1$ is generally one order of magnitude larger than $A_2$ and
$A_3$, which makes sense since the radial component is usually the
largest for nongravitational accelerations.

According to the properties of the comet population we considered
three different scenarios as described in Table~\ref{tab:scenarios}:
the ballistic scenario corresponds to JPL solution 46; the
``reference'' scenario uses typical values of the nongravitational
parameters; the ``wide'' scenario assumes extreme values of the
nongravitational parameters. We selected the $A_1$ uncertainty so that
its range would span 
from $0$ au/d$^2$ to twice the nominal value at $3\sigma$. For $A_2$
and $A_3$ the nominal value is $0$ au/d$^2$ since these components can be either
positive or negative, while $A_1$ can only be positive.

\begin{table}[ht]
  \caption{A priori values and 3$\sigma$ uncertainties of
    nongravitational parameters for the three scenarios.
\label{tab:scenarios}}
\begin{center}
\begin{tabular}{lccc}
\hline
Scenario & $A_1$ (au/d$^2$) & $A_2$ (au/d$^2$) & $A_3$ (au/d$^2$)\\
\hline
Ballistic & $0 \pm 0$  & $0 \pm 0$ & $0 \pm 0$\\
Reference & $(1 \pm 1) \times 10^{-8}$ & $(0 \pm 2) \times 10^{-9}$ & $(0 \pm 2) \times 10^{-9}$\\
Wide & $(1 \pm 1) \times 10^{-6}$ & $(0 \pm 2) \times 10^{-7}$ & $(0 \pm 2) \times 10^{-7}$\\ 
\hline
\end{tabular}
\end{center}
\end{table}

\begin{figure}[t]
\begin{center}
\includegraphics[width=5in]{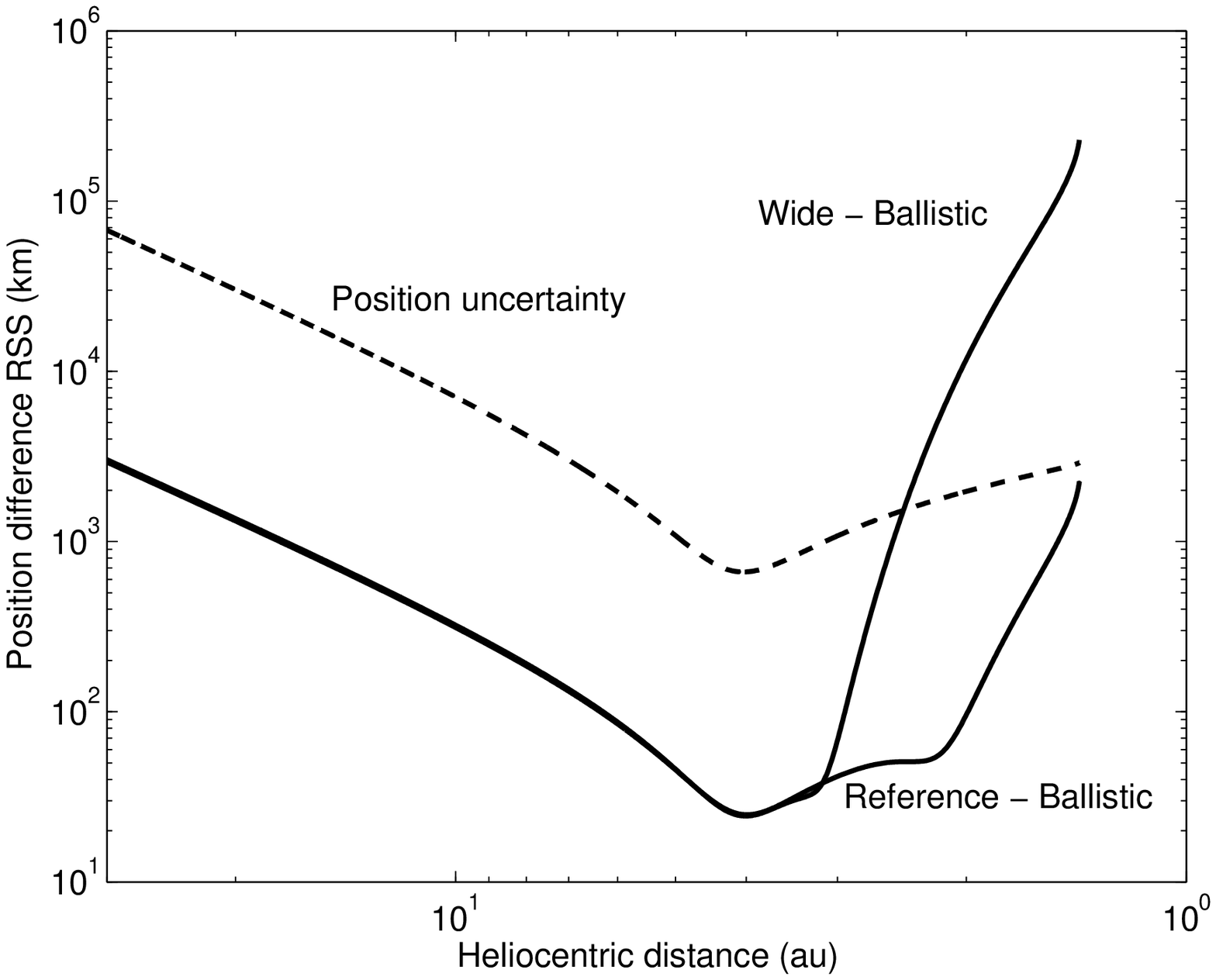}
\caption{Magnitude of the position difference between the reference
  and ballistic solutions, and between the wide and ballistic
  solutions, as a function of heliocentric distance. The dashed line
  is the semimajor axis of the $1 \sigma$ uncertainty ellipsoid of the
  ballistic solution.}
\label{fig:time_series}
\end{center}
\end{figure}

Figure~\ref{fig:time_series} shows the position difference among the
three scenarios compared to the position uncertainty of the ballistic
solution. The available observations put a strong constraint the
trajectory of C/2013~A1 for heliocentric distances between 3 au and 8
au from the Sun. Outside of this distance range we have no
observations and therefore the uncertainty increases. Because of the
fast decay of the $g(r)$ function in Eq.~\eqref{eq:nongravs} the
contribution of nongravitational accelerations for large heliocentric
distances is well within the uncertainty and so the trajectory of
C/2013~A1 in the past is not significantly affected. It is worth
pointing out that the function $g(r)$ represents water sublimation
while distant activity is not driven by water and therefore may be
inaccurate a large distances. However, for such large distances the
position uncertainty is large enough to make this possible discrepancy
irrelevant. Finally, for smaller heliocentric distances
nongravitational perturbations become relevant and can affect the
predictions for the Mars encounter, especially in the wide scenario.

\begin{figure}[t]
\begin{center}
\includegraphics[width=5in]{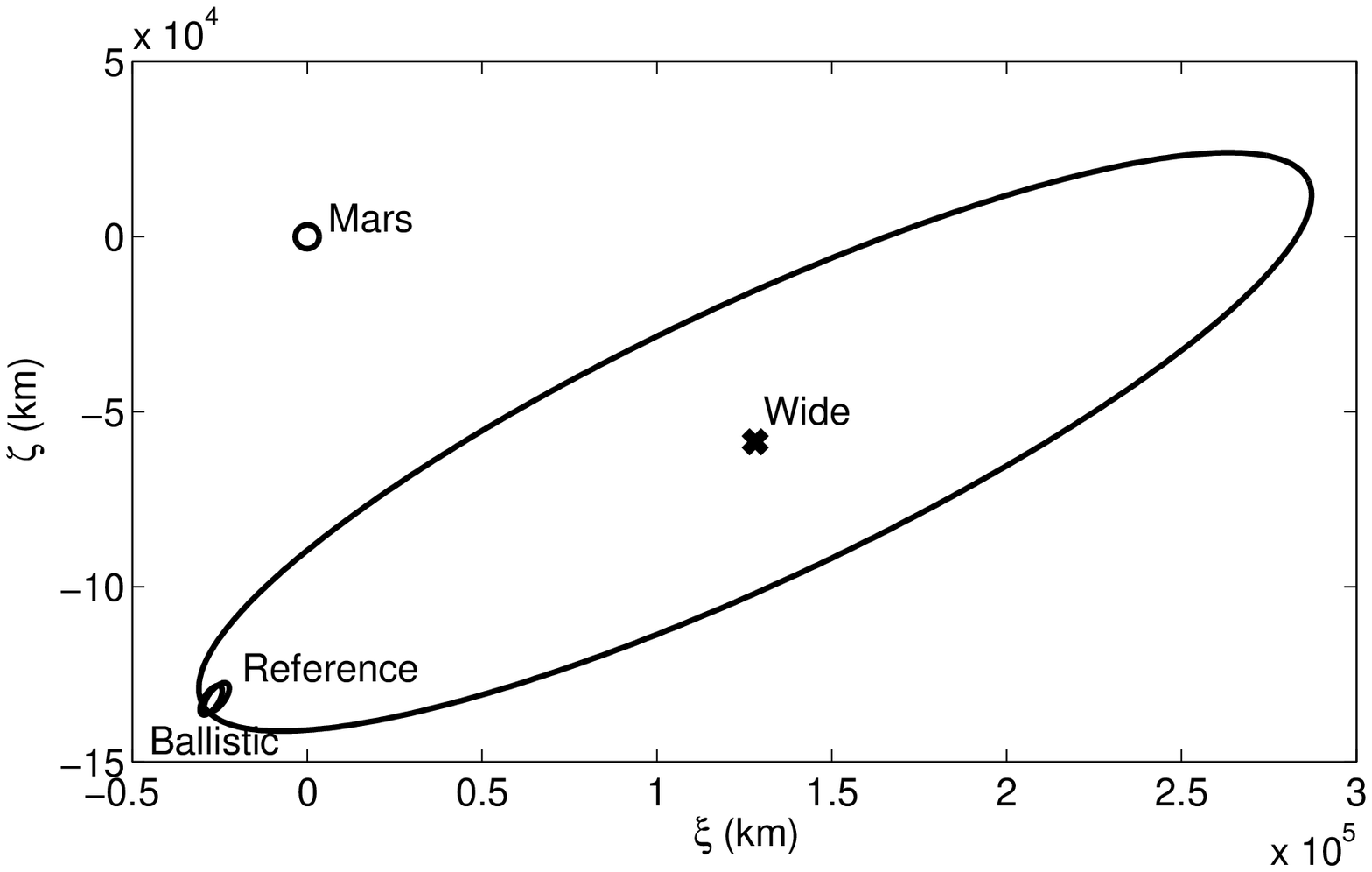}
\caption{Projection on the $b$-plane of C/2013~A1 3$\sigma$
  uncertainty according to different scenarios for nongravitational
  perturbations.  The ballistic and reference solutions are almost
  indistinguishable.}
\label{fig:ellipses}
\end{center}
\end{figure}

For the three different scenarios, Table~\ref{tab:b_plane} gives the
close approach information while Fig.~\ref{fig:ellipses} shows the
projection of the orbital uncertainties on the $b$-plane. The
ballistic and reference solutions provide very similar predictions,
from which we conclude that nongravitational perturbations will not
significantly affect the orbit unless they are larger than
expected. The wide solution, which has to be regarded as an extreme
case, produces a significantly different nominal prediction and quite
a large uncertainty.  In all three scenarios, the nominal close
approach distance is more than 130,000 km from Mars and therefore
there is no chance of an impact between the nucleus of C/2013~A1 and
Mars.

\begin{table}[ht]
\caption{Close approach parameters and uncertainties for the three
  scenarions. The table shows the $b$-plane coordinates, the
  semimajor axis of the 3$\sigma$ uncertainty projected on the
  $b$-plane, and the time of closest approach.
  \label{tab:b_plane}}
\begin{center}
\begin{tabular}{lcccc}
\hline
Scenario     & $\xi$ (km) & $\zeta$ (km) & 3$\sigma$ SMA (km) & TCA (TDB) $\pm$ 3$\sigma$ \\
\hline
Ballistic & -27,445 & -132,407  & 4789 & 2014 Oct 19 18:30 $\pm$ 3 min\\
Reference & -25,865 & -131,671 & 5047 & 2014 Oct 19 18:30 $\pm$ 3 min\\
Wide & 128,124 & -58,610 & 174,882 & 2014 Oct 19 19:15 $\pm$ 45 min\\
\hline
\end{tabular}
\end{center}
\end{table}

\section{Uncertainty evolution}
The predictions and the uncertainty provided so far are based on the
optical astrometry available as of March 15, 2014. At the time of
submission of this paper (April 2014), comet C/2013~A1 was difficult
to observe because of the low solar elongation.  On June 18, 2014 the
solar elongation becomes larger than 60$^\circ$ and we therefore
expect observations to resume, which will help in further constraining
the trajectory of C/2013~A1. To quantify the effect of future optical
astrometry, we simulated geocentric optical observations, with two
observations every five nights.

\begin{figure}[t]
\begin{center}
\includegraphics[width=5in]{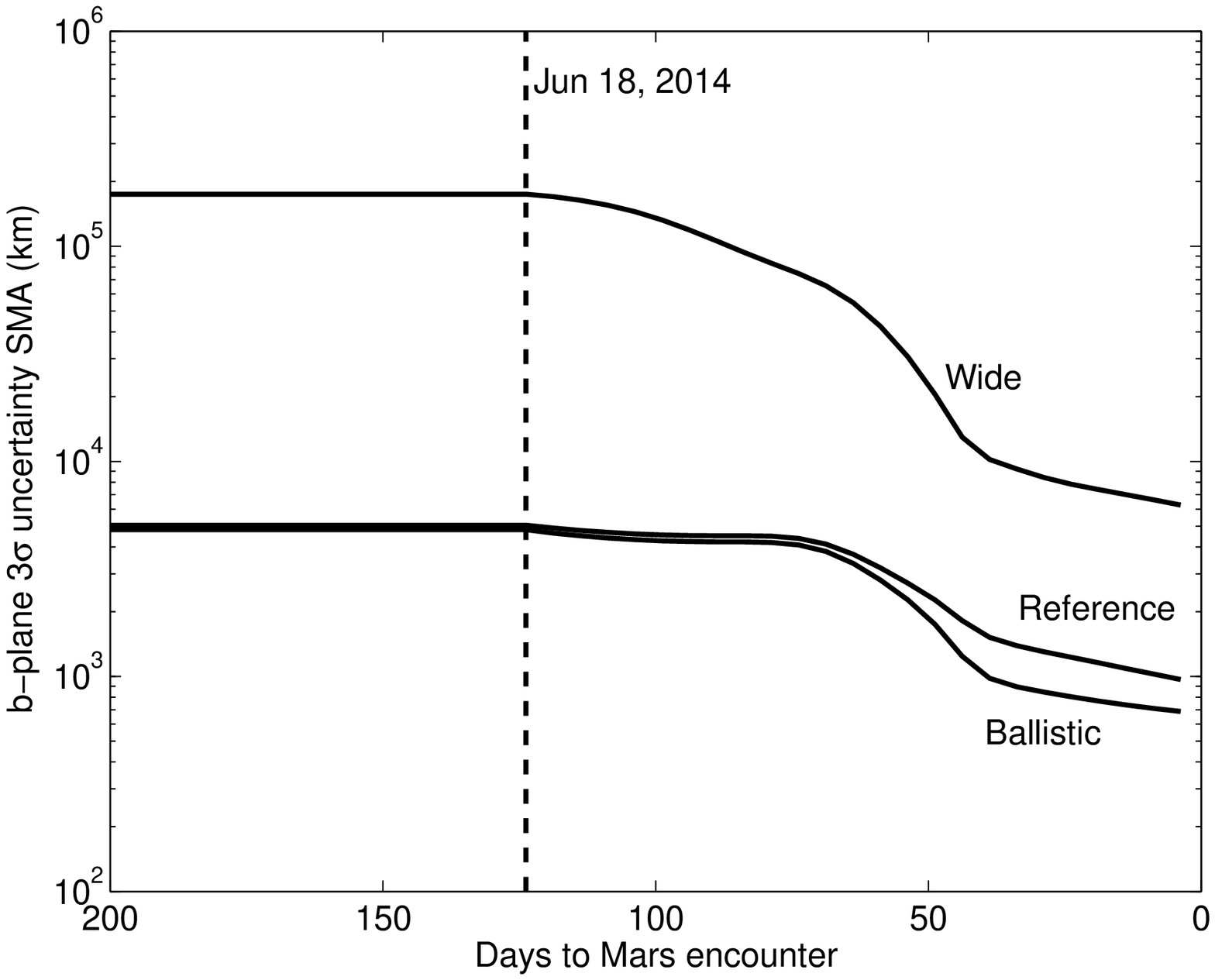}
\caption{Expected evolution of the $b$-plane position uncertainty. The
  curves represent the semimajor axis of the projection on the
  $b$-plane of the 3$\sigma$ uncertainty ellipse for the three
  scenarios. The vertical bar corresponds to Jun 18, 2014 when the
  solar elongation of C/2013~A1 becomes larger than 60$^\circ$.}
\label{fig:sma_ev}
\end{center}
\end{figure}

Figure~\ref{fig:sma_ev} shows the evolution of the position
uncertainty on the $b$-plane. The curves represent the semimajor axis
of the projection of the $3\sigma$ uncertainty ellipsoid on the
$b$-plane. The ballistic and reference solution curves are close, with
an uncertainty that goes from the current 5000 km to less than 1000 km
when all the pre-encounter observations are accounted for. The wide
solution has a much larger uncertainty that decreases to a minimum of
about 6000 km.

\begin{figure}[t]
\begin{center}
\includegraphics[width=5in]{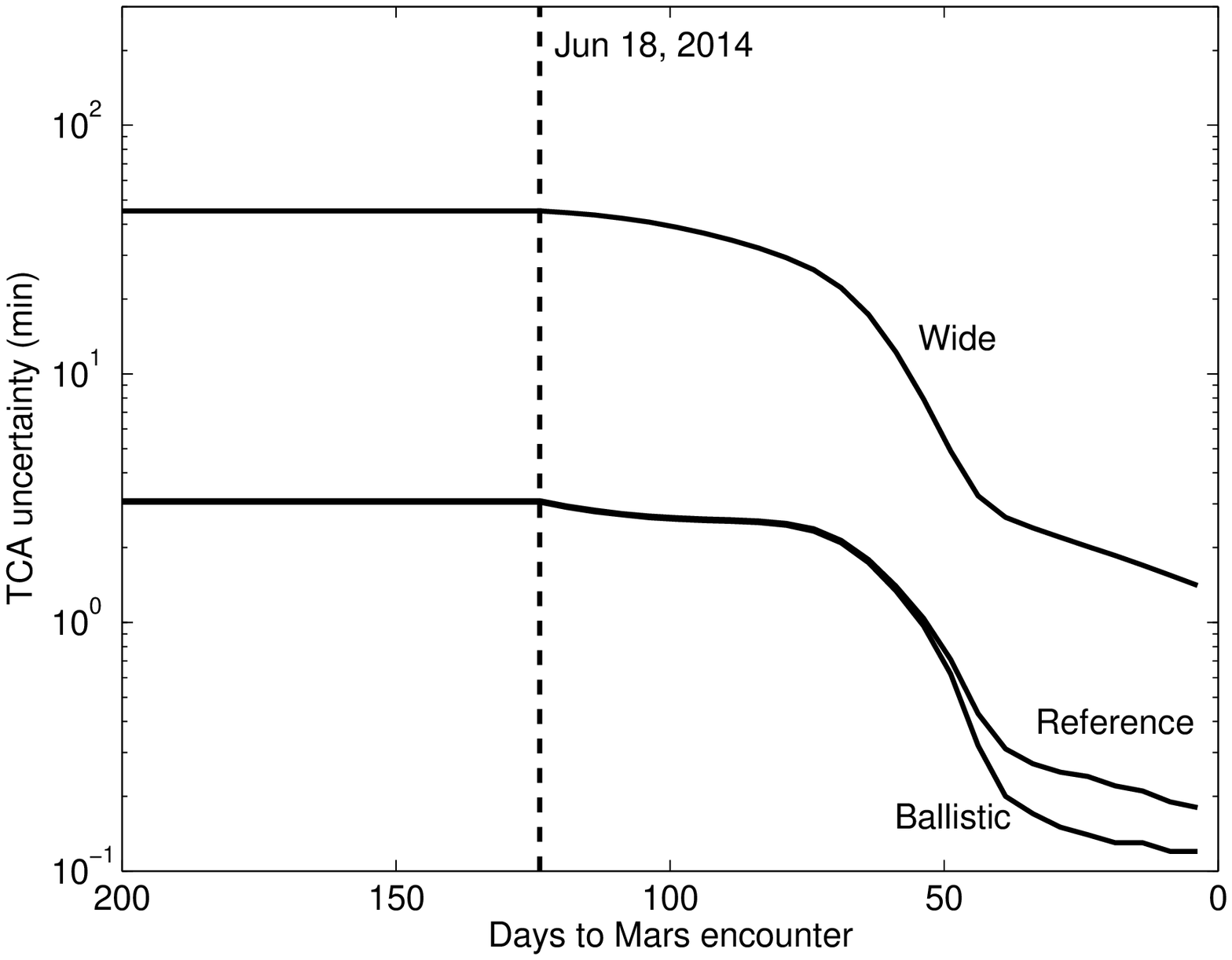}
\caption{Expected evolution of the 3$\sigma$ uncertainty of the
  closest approach epoch. The vertical bar corresponds to Jun 18, 2014
  when the solar elongation of C/2013~A1 becomes larger than
  60$^\circ$.}
\label{fig:tca_ev}
\end{center}
\end{figure}

Figure~\ref{fig:tca_ev} shows the $3\sigma$ uncertainty evolution for
the close approach epoch. The ballistic and reference scenarios have a
current uncertainty of 3 min and this uncertainty decreases to less
than 0.2 min right before the close approach. For the wide scenario
the uncertainty goes from 45 min down to 1--2 min.

\begin{figure}[t]
\begin{center}
\includegraphics[width=5in]{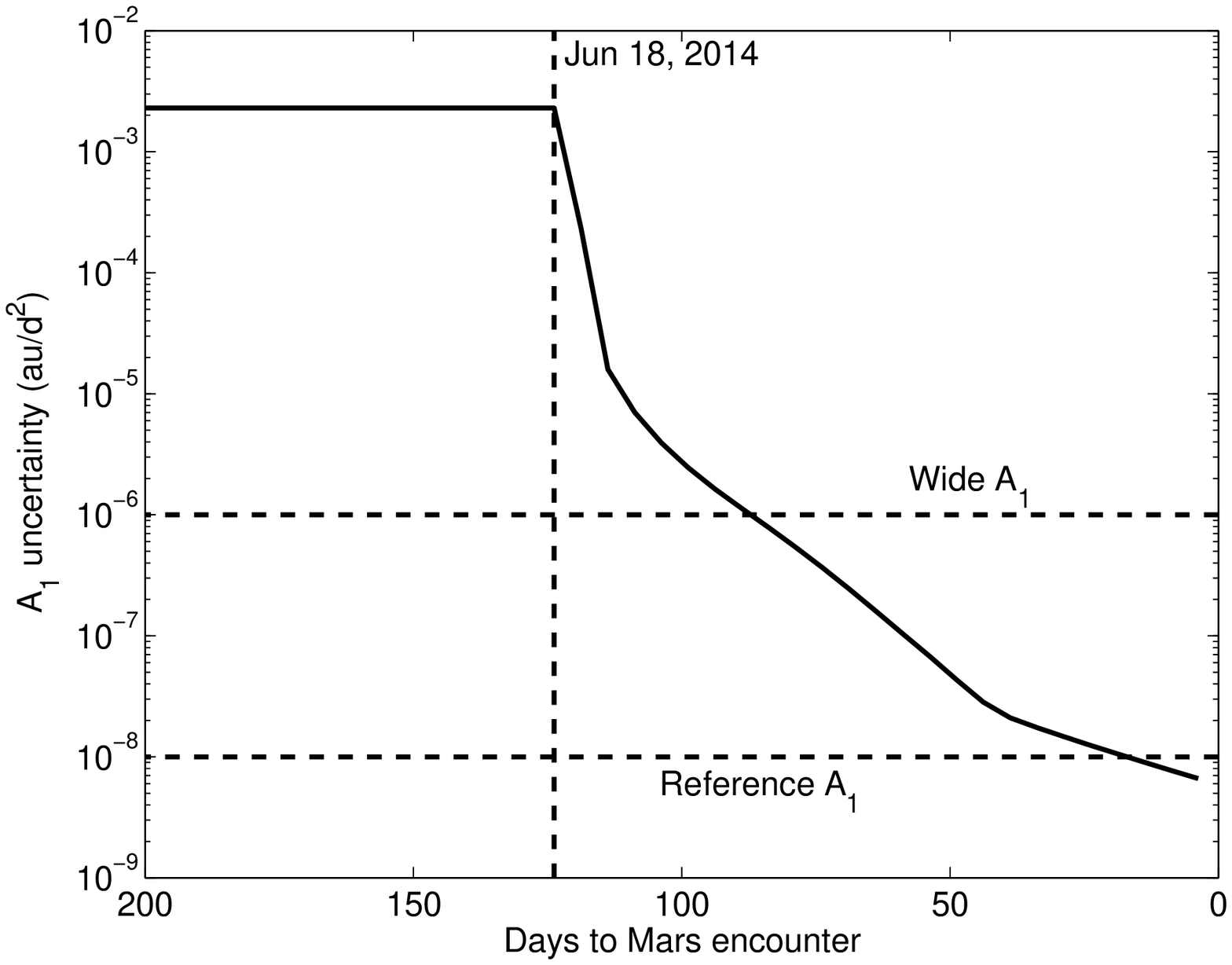}
\caption{Expected evolution of the $A_1$ uncertainty ($1\sigma$). The
  horizontal dashed lines are for the nominal values of $A_1$ in the
  reference and wide scenarios. The vertical bar corresponds to Jun
  18, 2014 when the solar elongation of C/2013~A1 becomes larger than
  60$^\circ$.}
\label{fig:ng_ev}
\end{center}
\end{figure}

As already discussed in Sec.~\ref{sec:nongravs}, the wide solution
produces predictions significantly different from the ballistic and
reference solutions. Thus, at some point observations will reveal
whether or not the nongravitational perturbations are behaving as in
the wide scenario. Figure~\ref{fig:ng_ev} shows the uncertainty in
$A_1$ when estimated from the orbital fit as a function of time. When
this uncertainty becomes smaller than a given value of $A_1$, the
observation dataset reveals such $A_1$ value if it is real. By
comparing the uncertainty evolution to the nominal values of $A_1$
assumed for the different scenarios, we can see that large
nongravitational accelerations to the level assumed in the wide
scenario are detectable about 90 days before the close encounter. On
the other hand, the reference solution becomes distinct from the
ballistic solution only a couple of weeks before the encounter.

Some skilled observers are capable of gathering comet observations
even for solar elongations smaller than 60$^\circ$. Therefore, we also
simulated observations using 40$^\circ$ as a lower threshold for the
solar elongation, which makes it possible to collect new observations
for C/2013~A1 starting on May 7, 2014. However, the improvement in the
uncertainties discussed above is a factor of 1.3 or less and is
therefore not relevant.

\section{Dust tail}
Though an impact the nucleus of C/2013~A1 on Mars is ruled out, there
is a chance that dust particles in the tail could reach Mars and some
of the orbiting spacecrafts. Due to their small size, the motion of
dust particles is strongly affected by solar radiation pressure. It is
therefore to use the $\beta$ parameter \citep{Burns79},
i.e., the non-dimensional number corresponding to the ratio between solar
radiation pressure and solar gravity. In terms of physical properties,
$\beta$ is proportional to the area-to-mass ratio and inversely
proportional to both the density and to the radius of the particle:
\begin{equation}
\beta = \frac{0.57\, Q}{a\rho}
\end{equation}
where $a$ is the particle radius in $\mu$m, $\rho$ is the density in
g/cc, and $Q$ is the solar radiation pressure efficiency coefficient. 

\begin{figure}[t]
\begin{center}
\includegraphics[width=5in]{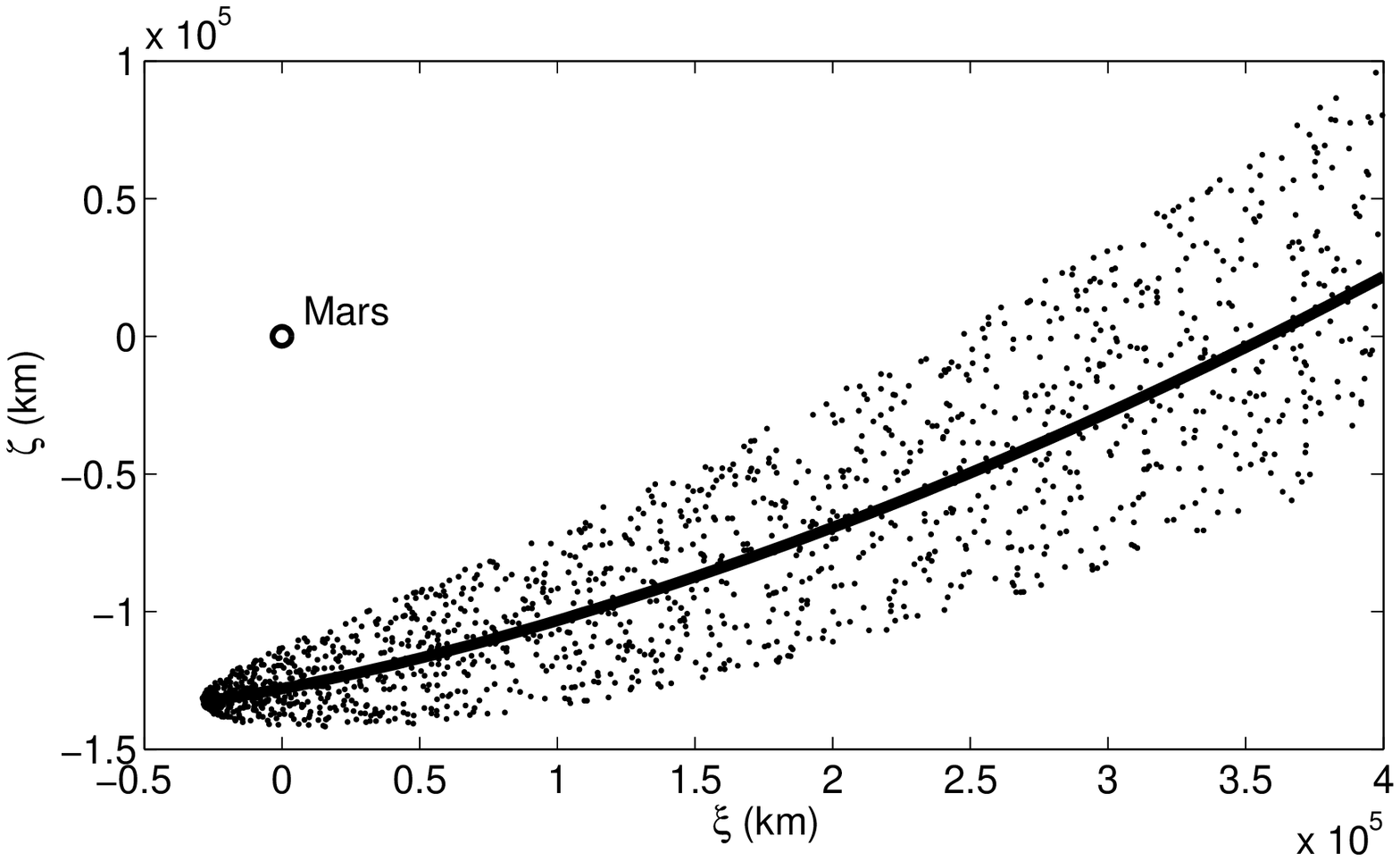}
\caption{Projection on the $b$-plane of particles ejected with $\Delta
  v = 10$ m/s and for $\beta = 0.01$. The solid line represents the
  position of the particles with no $\Delta v$.}
\label{fig:cloud}
\end{center}
\end{figure}

For each ejected particle, the location on the $b$-plane for the Mars
encounter is determined by the $\beta$ parameter, the heliocentric
distance $r$ at which the particle is ejected (or the ejection epoch), and the
ejection velocity $\Delta\mathbf v$. Figure~\ref{fig:cloud} shows the
typical behavior using as an example $\beta = 0.01$ and $\Delta v =
|\Delta\mathbf v| = 10$ m/s. For each given $\beta$ we have a curve on
the $b$-plane corresponding to zero ejection velocity. This curve can
be parameterized by the heliocentric distance at which the ejection
takes place. Finally, the ejection velocity $\Delta\mathbf v$ yields
dispersion around the curve: the larger the $\Delta v$ the wider
the dispersion.

The ejection velocity depends on the particle size and density, as
well as the heliocentric distance at which the particle is ejected
\citep{Whipple51}. Since cometary activity is very hard to predict,
modeling the ejection velocities is a complicated task and is subject
to continuous updates as additional observations are
available. Therefore, we decided to adopt a different approach: for
given ejection distance $r$ and $\beta$ parameter we computed
the minimum $\Delta v$ required to reach Mars. In mathematical terms
we look for the tridimensional $\Delta\mathbf v$ that is a minimum
point of $\Delta v^2 = |\Delta \mathbf v|^2$ under the constraint that
the particle reaches Mars, i.e., $(\xi,\zeta)(r, \beta, \Delta \mathbf
v) = (0, 0)$.

This problem is a typical example of finding the minima of a function
subject to equality constraints. Thus, we can solve this problem by
means of the Lagrange multipliers, i.e., the $\Delta\mathbf v$ we are looking
for must satisfy the following system of equations:
\begin{equation}\label{eq:lagrange}
\begin{dcases}
(\xi, \zeta)(r, \beta, \Delta\mathbf v) = (0, 0)\\
\frac{\partial |\Delta \mathbf v|^2}{\partial \Delta\mathbf v} =
\lambda_1 \frac{\partial \xi}{\partial \Delta\mathbf v}(r, \beta, \Delta
\mathbf v) +
\lambda_2 \frac{\partial \zeta}{\partial \Delta\mathbf v}(r, \beta,
\Delta \mathbf v)
\end{dcases}
\end{equation}
where $\lambda_1$ and $\lambda_2$ are free parameters. To solve this
system, we first tested the linearity of $(\xi, \zeta)$ in
$\Delta\mathbf v$ and then linearized system~\eqref{eq:lagrange}
around $\Delta\mathbf v =0$, thus obtaining the following linear
system:
\begin{equation}\label{eq:lagrange_lin}
\begin{dcases}
(\xi, \zeta)(r, \beta, \Delta\mathbf v) =(\xi,\zeta)(r, \beta, 0) +
\frac{\partial (\xi, \zeta)}{\Delta
  \mathbf v}(r, \beta, 0) \Delta \mathbf v = (0, 0)\\
2\Delta\mathbf v = \lambda_1 \frac{\partial \xi}{\partial \Delta\mathbf v}(r, \beta, 0) +
\lambda_2 \frac{\partial \zeta}{\partial \Delta\mathbf v}(r, \beta,
0).
\end{dcases}
\end{equation}
To compute the required $\Delta\mathbf v$, we followed these steps:
\begin{itemize}
\item We sampled $\beta$ in $\log$-scale from $10^{-6}$ to 1 and $r$
  from 1.4 au to 30 au;
\item For each couple $(r, \beta)$ we computed the $b$-plane
  coordinates $(\xi, \zeta)$ obtained without ejection velocity as
  well as a finite difference approximation of the $(\xi, \zeta)$
  partials with respect to $\Delta\mathbf v$;
\item We solved system~\eqref{eq:lagrange_lin}.
\end{itemize}
We scaled the resulting $\Delta\mathbf v$ to account for the size of
Mars and the 3$\sigma$ uncertainty of the particle projection on
the $b$-plane. For this analysis we used the ballistic solution as
reference trajectory.

\begin{figure}[t]
\begin{center}
\includegraphics[width=5in]{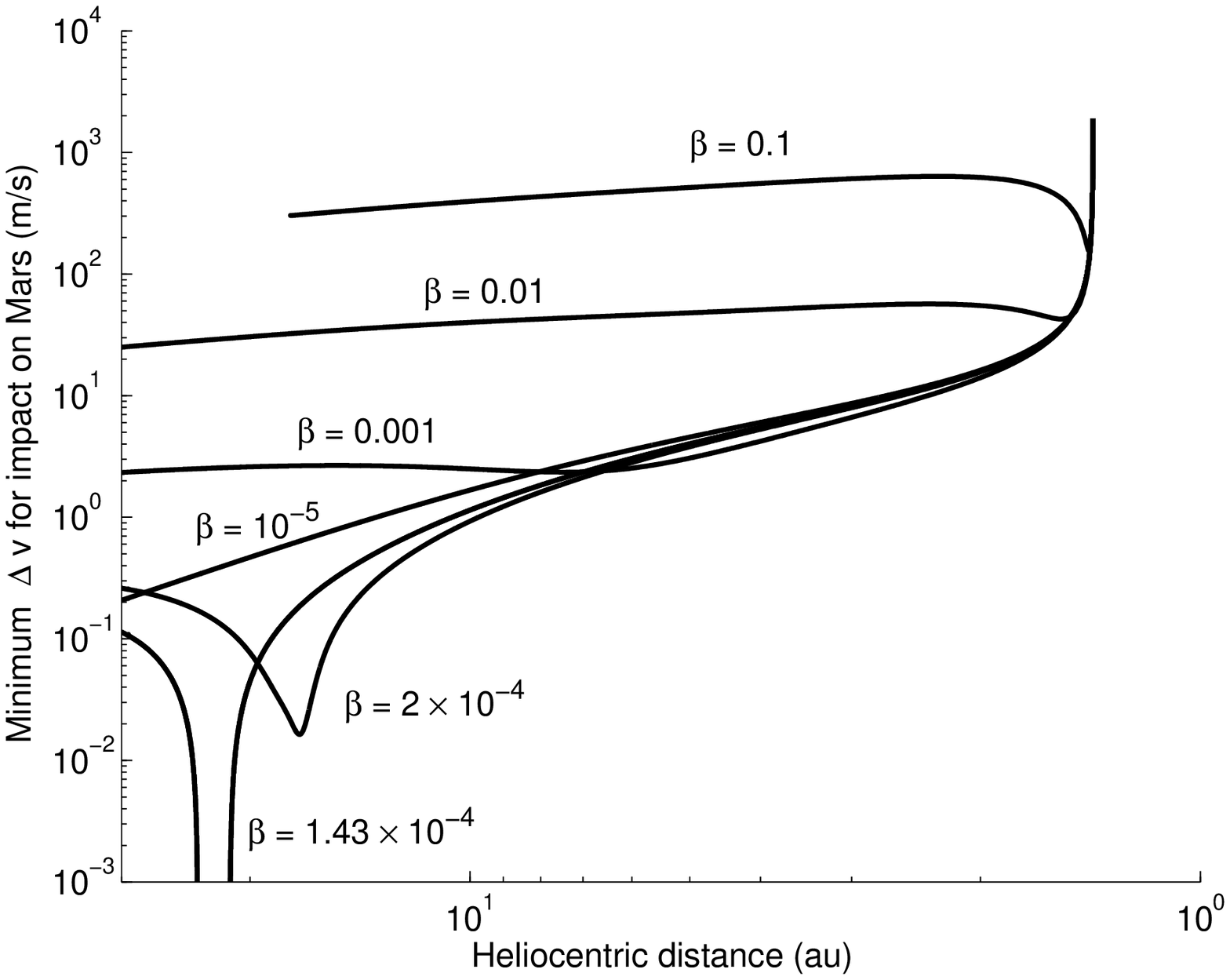}
\caption{For different values of $\beta$, required $\Delta v$ to reach
  Mars as a function of the heliocentric distance at which the
  ejection takes place.}
\label{fig:reqdv}
\end{center}
\end{figure}

Figure~\ref{fig:reqdv} shows the required $\Delta v$ needed to reach
Mars as a function of the heliocentric distance at which the ejection
takes place for different values of $\beta$. On the right side of the
plot the required velocities are almost the same. This behavior makes
sense as the closer we get to Mars the less time is available for
solar radiation pressure to affect the trajectory. Therefore, the
required ejection velocity is almost independent of the particle size
and density. For $\beta = 1.43 \times 10^{-4}$ we can see that the
required velocity goes to zero for heliocentric distances around 22.5
au. The reason for this is that the curve on the $b$-plane defined by
this particular value of $\beta$ passes through the center of
Mars. Thus, if ejected at the right distance, i.e., 22.5 au, the
particle reaches Mars under the action of solar radiation pressure,
with no ejection velocity at all. It is also worth noticing that the
$\beta = 0.1$ curve does not go all the way back to 30 au because, for
such a high $\beta$, solar radiation pressure is extremely strong and
the particle does not even experience the close encounter with Mars if
ejected too far in advance.

\begin{figure}[t]
\begin{center}
\includegraphics[width=5in]{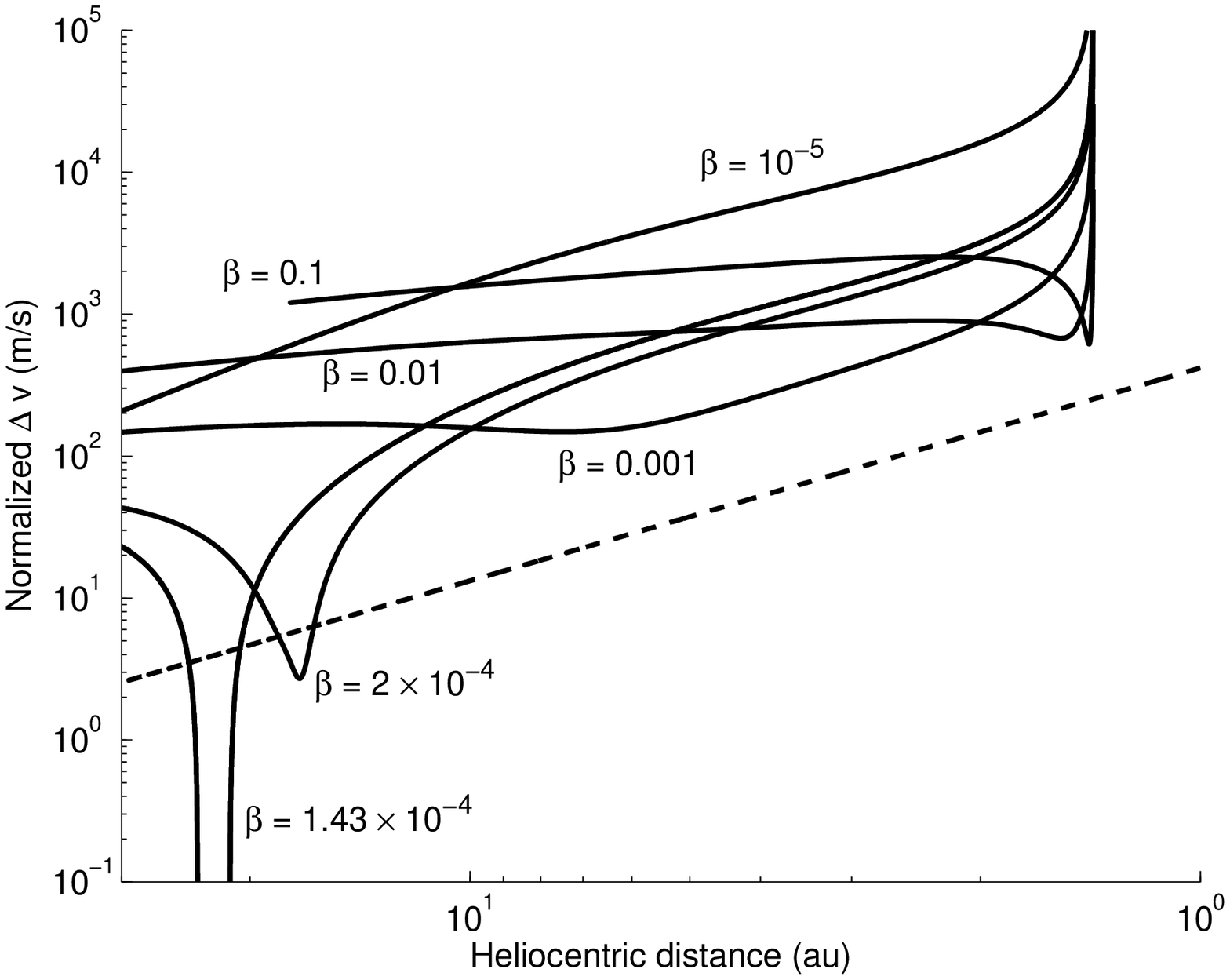}
\caption{Required $\Delta v$ to reach Mars multiplied by
  $(1/\beta)^{0.6}$. The dashed line corresponds to $\Delta v = 418
  \mbox{ m/s } \beta^{0.6} (1 \mbox{ au }/r)^{1.5}$. Impacts are
  possible only for particles ejected more than 16 au from the Sun and
  with $\beta \sim 2 \times 10^{-4}$ or smaller.}
\label{fig:farnham}
\end{center}
\end{figure}

The results obtained so far can be used to assess the possibility that
particles of a given size could reach Mars for a given ejection
velocity model. For instance, the best fit for the ejection velocity
according to \citet{Farnham14} is
\begin{equation}\label{eq:farnham}
  \Delta v = 418\mbox{ m/s } \left(\frac{\beta}{1}\right)^{0.6}\left(\frac{1\mbox{
        au }}{r}\right)^{1.5}.
\end{equation}
As shown in Fig.~\ref{fig:farnham}, we can scale the required velocity
to $\beta = 1$ and make a comparison to the velocity given by
\eqref{eq:farnham}. We can see that, according to this ejection
velocity model, impacts are possible only for particles with $\beta
\sim 2 \times 10^{-4}$ or smaller ejected at more than $\sim 16$ au
from the Sun.

\begin{figure}[t]
\begin{center}
\includegraphics[width=5in]{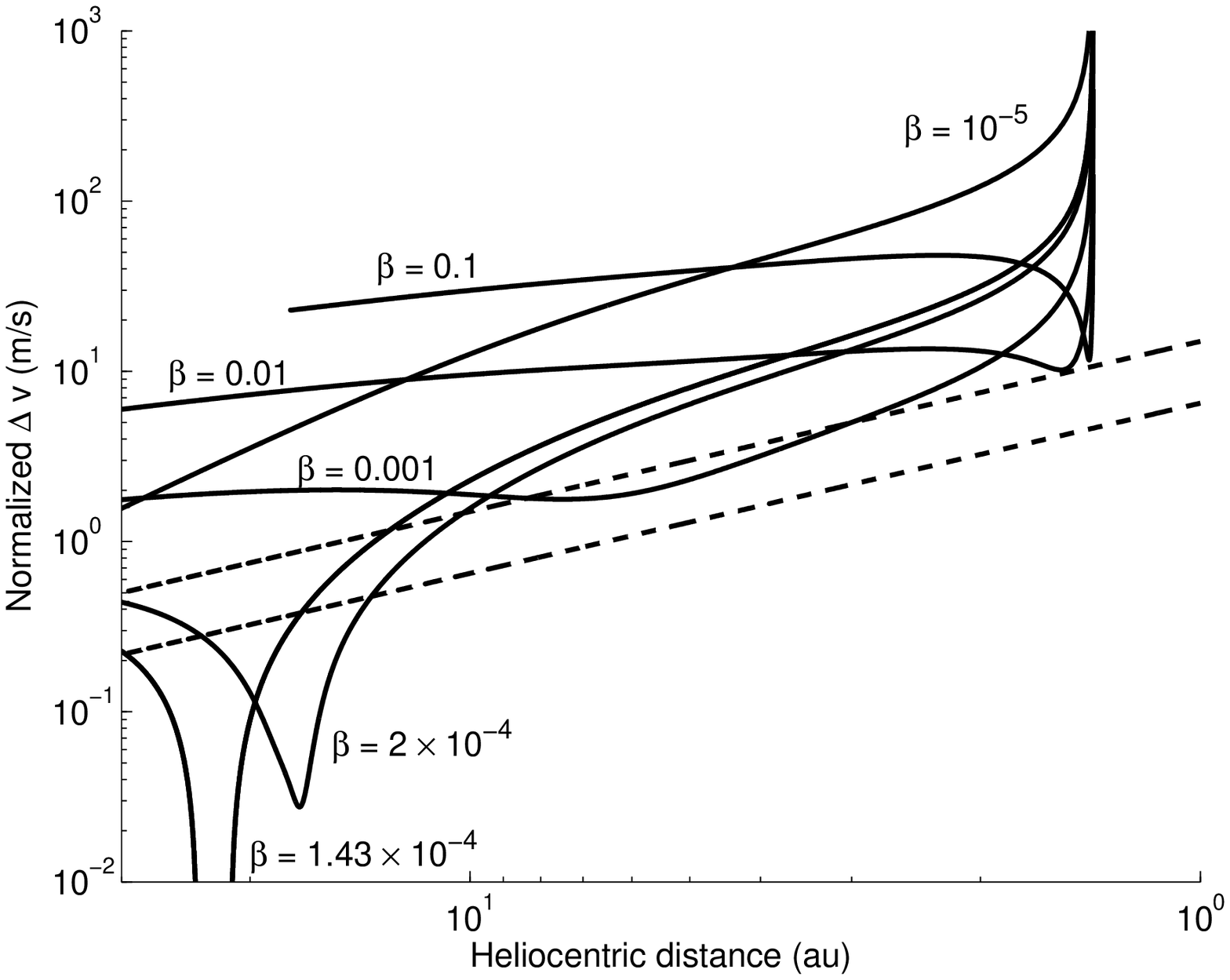}
\caption{Required $\Delta v$ to reach Mars multiplied by $(5.7 \times
  10^{-4}/\beta)^{0.5}$. The lower dashed line corresponds to $\Delta
  v = 1.3 \mbox{ m/s } (\beta/5.7 \times 10^{-4})^{0.5} (5 \mbox{ au
  }/r)^{1}$. In this case impacts are possible for particles ejected
  more than 13 au from the Sun and $\beta \sim 2 \times 10^{-4}$ or
  smaller. The upper dashed line corresponds to $\Delta v = 3 \mbox{
    m/s } (\beta/5.7 \times 10^{-4})^{0.5} (5 \mbox{ au }/r)^{1}$. In
  this case impacts are possible also for $\beta = 0.001$ and particles
  ejected as close as $\sim 3$ au from the Sun.}
\label{fig:tricarico}
\end{center}
\end{figure}

Figure~\ref{fig:tricarico} shows a comparison to the ejection velocity
model considered by \citet{Tricarico14}:
\begin{equation}\label{eq:tricarico}
\Delta v = 1.3 \mbox{ m/s } \left(\frac{\beta}{5.7 \times 10^{-4}}\right)^{0.5}\left(\frac{5\mbox{
      au }}{r}\right)^{1}.
\end{equation}
In this case impacts are possible only for particles ejected more than
13 au from the Sun and $\beta \sim 10^{-4}$. The figure also makes the
comparison for larger ejection velocities \citep[also considered by][]{Tricarico14}:
\begin{equation}\label{eq:tricarico2}
\Delta v = 3 \mbox{ m/s } \left(\frac{\beta}{5.7 \times 10^{-4}}\right)^{0.5}\left(\frac{5\mbox{
      au }}{r}\right)^{1}.
\end{equation}
In this case impacts are possible also for $\beta = 0.001$ and
particles ejected as close as $\sim 3$ au from the Sun.

We conclude that impacts are possible only in one of these two unlikely
cases:
\begin{itemize}
\item millimeter to centimeter dust grains are ejected from the
  nucleus more than 13 au from the Sun;
\item the ejection velocities are larger than current estimates by a
  factor $>2$. 
\end{itemize}
In the first case the particles can reach Mars during a 20 min
interval centered at the time that Mars crosses the orbit of
C/2013~A1, i.e., Oct 19, 2014 at 20:09 (TDB). In the second case the
time interval is wider and goes from 43 min to 130 min after the
nominal close approach of the ballistic trajectory. For an analysis of
the probability distribution of the arrival times, see
\citet{Tricarico14} and \citet{Kelley14}.


\section{Conclusions}
To study the Oct 19, 2014 encounter with Mars, we analyzed the
trajectory of comet C/2013~A1 (Siding Spring). The ballistic orbit has
a closest approach with Mars at 135,000 km $\pm$ 5000 km at 18:30
TDB.

Nongravitational perturbations are not yet detectable for C/2013~A1,
so we assumed known nongravitational parameters for known comets in
the catalog. In case of typical nongravitational perturbations there
are no relevant differences from the ballistic trajectory. On the
other hand, unexpectedly large nongravitational accelerations would
produce significant deviations that should become detectable in the
observation dataset by the end of July 2014.  However, even in the
case of unexpectedly large nongravitational perturbations, the nucleus
C/2013~A1 cannot reach Mars.

To analyze the risk posed by dust grains in the tail, we computed the
required ejection velocities as a function of the heliocentric
distance at which the particle is ejected and the particle's $\beta$
parameter, i.e., the ratio between solar radiation pressure and solar
gravity.  By comparing our results to the most updated modeling of
dust grain ejection velocities, impacts are possible only for $\beta$
of the order of $10^{-4}$, which, for a density of 1 g/cc, corresponds
to millimeter to centimeter particles. However, the particles have to
be ejected at more than 13 au, which is generally considered
unlikely. See \citet{Kelley14} for a discussion of the maximum
liftable grain size at these distances. The arrival times of these
particles are in an interval of about 20 minutes around the time that
Mars crosses the orbit of C/2013~A1, i.e., Oct 19, 2014 at 20:09
TDB. In the unlikely case that ejection velocities are larger than
currently estimated by a factor $>2$, impacts are possible for
particles with $\beta = 0.001$ that are ejected as close as $\sim 3$
au from the Sun. These impacts would take place from 43 min to 130 min
after the nominal ballistic close approach of the nucleus.

As the comet gets closer to the inner solar system, new observations
will be available and will allow better constraints on the dust grain
ejection velocity profile. Our analysis can be used as a reference to
quickly figure out what particles can reach Mars and the heliocentric
distance at which they would have to have been ejected. In the
unlikely case that future astrometry reveals unexpectedly large
nongravitational perturbations, the required velocity to reach Mars
for particles ejected within 2 au from the Sun can change and the
presented analysis will need to be refined.

\section*{Acknowledgments}

\noindent Part of this research was conducted at the Jet Propulsion
Laboratory, California Institute of Technology, under a contract with
the National Aeronautics and Space Administration.

\noindent Copyright 2014, California Institute of
Technology.

\bibliography{ref}
\bibliographystyle{apalike}

\end{document}